\documentstyle[12pt,psfig]{article}

\newcommand{\kms}{\mbox{km~s$^{-1}$}}
\newcommand{\mols}{\mbox{molec.~s$^{-1}$~}}

\newcommand{\permil}{\mbox{$^0\!/\!_{00}$}}
\renewcommand{\deg}{\mbox{$^{\circ}$}}

\begin{document}

\title{Submillimetre observations of comets with Odin: 2001--2005\footnote{
	Odin is a Swedish-led satellite project funded jointly 
	by the Swedish National Space Board (SNSB), the Canadian Space Agency (CSA), 
	the National Technology Agency of Finland (Tekes) and the
	Centre National d'\'Etudes Spatiales (CNES, France). 
	The Swedish Space Corporation is the
	prime contractor, also responsible for Odin operations.}}

\author{Nicolas Biver$^1$\footnote{Corresponding author: tel. 33 1 45077809, 
Fax: 33 1 45077939, e-mail: nicolas.biver@obspm.fr}, 
Dominique Bockel\'ee-Morvan$^1$, Jacques Crovisier$^1$, Alain Lecacheux$^1$,\\
        Urban Frisk$^2$, {\AA}ke Hjalmarson$^3$, Michael Olberg$^3$,\\
        Hans-Gustav Flor\'en$^4$, Aage Sandqvist$^4$ and Sun Kwok$^{5,6}$}
\date{\today}
\maketitle
\noindent
{\small $^1$ LESIA, CNRS UMR 8109, Observatoire de Paris, 5 pl. Jules Janssen, F-92190 Meudon, France}\\
{\small $^2$ Swedish Space Corporation, PO Box 4207, SE-17104 Solna, Sweden}\\
{\small $^3$ Onsala Space Observatory, SE-43992 Onsala, Sweden}\\
{\small $^4$ Stockholm Observatory, SCFAB-AlbaNova, SE-10691 Stockholm, Sweden}\\
{\small $^5$ Dept. of Physics and Astronomy, University of Calgary, Calgary, AB T2N 1N4, Canada}\\
{\small $^6$ Inst. of Astron. \& Astrophys., Academia Sinica, PO Box 23-141, Taipei 106, Taiwan}\\

\abstract{The Odin satellite, launched in Feb.  2001, is equipped with
a 1.1-m submillimetre telescope.  Odin was used to observe the 557~GHz
line of water with high spectral resolution in 12 comets between 2001
and 2005.  Line shapes and spatial mapping provide information on the
anisotropy of the outgassing and constraints on water excitation,
enabling accurate measurements of the water production rate.  Five
comets were regularly observed over periods of more than one month to
monitor the variation of their water outgassing rate with heliocentric
distance.  Observing campaigns have been generally coordinated with
ground-based observations of molecular lines at Nan\c{c}ay, CSO or
IRAM~30-m telescopes to obtain molecular abundances relative to water.

Thanks to Odin's frequency coverage, it was also possible to detect
the H$_2^{18}$O 548~GHz line, first in comet 153P/Ikeya-Zhang in April
2002 (Lecacheux et al.  (2003)) and then in comets C/2002~T7 (LINEAR),
C/2001~Q4 (NEAT) and C/2004~Q2 (Machholz).  The $^{16}$O/$^{18}$O
isotopic ratio ($\approx450$) is consistent with the terrestrial
value.  Ammonia has been searched for in three comets through its
$J_K=1_0-0_0$ line at 572 GHz and was tentatively detected in
C/2001~Q4 and C/2002~T7.  The derived abundances of NH$_3$ relative to
water are 0.5\% and 0.3\%, respectively, similar to values obtained in
other comets with different techniques.  } \\


\section{Introduction}
	Water is the main constituent of the ices of cometary nuclei.
The study of cometary water is thus crucial for cometary science.
Measurements of water production rates allow us
to determine the relative abundances of cometary volatiles.
Several molecules can be observed in the ultraviolet, infrared and
radio domains, but the opacity of the 
Earth's atmosphere precludes the observation of water from the ground,
except for weak lines arising from highly excited rovibrational states.

	The Odin satellite (Nordh et al. 2003) was launched on 20 
February 2001 on a Sun synchronous polar orbit. Odin houses a radiometer
with a 1.1-m primary mirror and equipped with 5 receivers at 119~GHz and 
covering the 
486--504~GHz and 541--580~GHz bands that are in large part 
unobservable from the ground. Half of the time is dedicated to 
astronomical studies and the other half to aeronomical investigations. 
The main astronomical objectives are to search for O$_2$ and H$_2$O 
isotopologues emission in the Universe, from the Solar System to galaxies. 

	The first observation of the H$_2$O ($1_{10}-1_{01}$) fundamental
line of water at 556.936~GHz in a comet was obtained by the Submillimeter 
Wave Astronomical Satellite (SWAS, Neufeld et al. 2000) in 1999.
	Comets have been a major observing topic for Odin: the 
first results were reported by Lecacheux et al. (2003) and more recent 
observations by Hjalmarson et al. (2005). At the end of 2005, 
water has been detected
in 11 comets with Odin and H$_2^{18}$O and NH$_3$ lines were also observed. 
The 557~GHz water line is one the strongest cometary 
submillimetre lines. Its peak intensity is about 10~K in bright 
comets, comparable to the other brightest
lines in the sky from the Orion molecular cloud (Hjalmarson et al. 2003).

\section{Observations}

\subsection{Odin in-flight performances}

Odin is well suited for comet studies: it is equipped with single side-band 
receivers with system temperatures of 3000--3500 K and 
two to three receivers can be used simultaneously (Frisk et al. 2003).
When not in ``eclipse period'' (roughly between mid-May and mid-August
when part of Odin's orbit lies in the shadow of the Earth and power is 
limited), Odin can run three receivers simultaneously: the first two 
commonly used (``555B2'' and ``549A1'') are covering the band of the 
three water isotopologues (H$_2^{16}$O, H$_2^{17}$O and H$_2^{18}$O), 
and a third one ``572B1''
can be used to observe the fundamental line of NH$_3$ at 572.498~GHz.
Odin is equipped with three spectrometers: a low-resolution 
(1~MHz) wide band (1~GHz) acousto-optical spectrometer (AOS) and two 
high-resolution (177~kHz, reduced to 202~kHz after 2002) autocorrelators. 
High resolution (corresponding to $\approx$100 m~s$^{-1}$ at 557~GHz) is 
essential to resolve cometary lines, and particularly to study the
asymmetry of the 557~GHz water line in cometary atmospheres 
due to its optical thickness and self-absorption in 
the foreground coma (cf Lecacheux et al. (2003) and Section 2.2).
The Odin beam at 557~GHz is 2.2' and the main beam efficiency is about 0.85.

	Astronomical observations are usually divided into periods of 
about 60~min of pointing on the target per 96~min orbit. 
Achieving good pointing has been an area of concern with Odin. The 
first comet observed (C/2001~A2 (LINEAR)) was actually a bright line source 
used to calibrate the pointing of the satellite. The goal was to measure 
 precisely the offset between the telescope axis and the reference of the 
satellite. The pointing has been then regularly checked on bright 
sources like the compact Orion-KL H$_2$O outflow and the Jupiter 
continuum (Frisk et al. 2003). The offset
between the actual pointing and the telescope beam direction, estimated 
from the reconstructed attitude of the spacecraft, is now below 20". 
When two star-trackers are operational
the pointing accuracy can be better than 5", but typically at the beginning
or the end of an orbit, for $\sim$10~min, pointing and attitude reconstruction
rely on only one star-tracker and here the pointing error can be as large 
as 1', as verified by our mapping of some bright comets. Finally the three 
receivers, especially the ``555B2'' and ``549A1'' used for H$_2^{16}$O 
and H$_2^{18}$O observations, are slightly misaligned, by 15" at most.

Since 28.6 April 2003, i.e. the last orbit of comet 
153P/Ikeya-Zhang observations, the Odin observing mode of continuous 
tracking for moving targets is used for all comet observations. 
During the first two years, the standard comet observing mode was simply based
on a fixed pointing with coordinates updated regularly. This was done 
every 5--15 min, during which the comet was slightly drifting 
through the Odin beam because of its proper motion.

For all comet observations, data have been reduced using the reconstructed
attitude and latest orbital elements in order to compute positional offsets,
and make proper summations.
A few observations (concerning C/2000~WM$_1$ and C/2001~Q4) have been 
affected by pointing errors of up to 60" due to the lack of precision of the 
ephemerides at the time of the observations, but the loss of signal had little
consequences.
Finally, the choice of targets and viewing opportunities are limited by
the solar elongation constraint of Odin: 60\deg~ to 120\deg.

\subsection{Comets observed with Odin}
	We choose to mainly target the comets that were potentially active
enough for in-depth chemical investigations from ground-based and space 
observatories. Several comets
came close enough to the Earth ($\leq$0.4 AU) with a large enough outgassing
rate to undertake more detailed studies (Sections 3--6). The first four comets
observed have been briefly presented in Lecacheux et al. (2003). 
The detailed log of all observations and their analysis will be presented 
in a future paper.
	When possible, the Odin observations were coordinated with other 
radio observations: with OH observations at 18~cm with the Nan\c{c}ay 
radio telescope for the long-term monitoring of the water production rate 
and for assessing water and OH modelling (Colom et al. 2004);
with millimetre and submillimetre molecular observations with the Caltech 
Submillimeter Observatory (CSO) 10-m and the Institut de Radioastronomie 
Millim\'etrique (IRAM) 30-m observatories (Biver et al. 2006a), to obtain 
molecular abundances relative to water.

\begin{center} [Figure 1] \end{center}

\subsubsection{C/2001~A2 (LINEAR)}
	Comet C/2001~A2 (LINEAR) was initially a faint comet that underwent
a large brightness outburst (5 magnitudes) around 28 March 2001. 
It then kept a sustained activity for the following months and was 
extensively investigated from the ground (Biver et al. 2006a). On 27 April, 
3 months after launch, Odin made one of its first astronomical observations 
towards this comet: the spectrum corresponding to the 26~min of good 
pointing (within 40'') is shown in Fig.~\ref{h2oa22pk49p}.
Perihelion took place on 24 May 2001 at 0.78 AU.
The comet was then extensively mapped between 20 June and 9 July around 
the time when it was the closest to the Earth at 0.24 AU. 18 maps of
25-point grids $4\times4$' or $2\times2$' were planned using 90 Odin orbits.
On 27 April, pointing was still in early commissioning phase and only 
43~min integration out of 3 orbits yielded a detection of the comet.
Before 28 June, Odin timing was wrong and only about 10\% of data were useful.
Between 28 June and 9 July, still only 60\% of the orbits were correctly 
pointed to map the comet, but data were good enough to provide valuable
information on the telescope pointing.

\subsection{19P/Borrelly}
	This comet was observed shortly after the end of commissioning of
Odin, which still resulted in some mispointed orbits. This 6.8-year periodic
comet returned to perihelion on 15 September 2001 at 1.36 AU and was
flown over by Deep Space 1 on the 21st of September. Data and results have been
presented in detail in Bockel\'ee-Morvan et al. (2004).

\subsection{C/2000~WM$_1$ (LINEAR)}
	This new Oort comet had its perigee on 2 December 2001 at 0.32 AU.
Perihelion took place on 23 January 2002 at 0.56 AU but it was then more 
difficult to observe, being further away from the Earth and too close to 
the Sun for Odin (elongation 30\deg~ to 60\deg~ between 24 December 2001
and 11 March 2002). The first Odin observations took place on 7 and 8 December.
Observations at CSO were performed nearly in parallel, 
from 3 to 8 Dec. (Biver et al. 2006a). The second set of data was obtained on 
12 March 2002. Ephemerides were off by 60'' at the time of December 
observations, but maps were obtained and the signal was strong enough 
to detect the water line.  
The comet had a large outburst at the end of January that was followed 
by a sustained activity. Pre/post perihelion asymmetry in water production rate
is suggested from the Odin observations (Table~\ref{tabqp}).

\subsection{153P/Ikeya-Zhang}
	Comet 153P/2002~C1 (Ikeya-Zhang), discovered in February 2002, was 
found to be the return of the comet observed by Hevelius in 1661. 
It reached perihelion on 18 March 2002 at 0.51 AU and was an easy naked
eye comet. Its peak total outgassing rate neared $10^{30}$~\mols.
Perigee took place on 29 April 2002 at 0.40 AU. The comet was extensively 
studied at radio wavelengths during the March--May period (Biver et al. 2006a).
	Most results of the Odin observations, including the detection of 
H$_2^{18}$O, were presented in Lecacheux et al. (2003). Daily production 
rates are given in Table~\ref{tabqp}. The observed and simulated radial 
evolution of the H$_2$O line are illustrated in Fig.~\ref{profiliz}.

\begin{center} [Figure 2] \end{center}

\subsection{C/2002~X5 (Kudo-Fujikawa)}
	This comet passed its perihelion at only 0.19 AU 
from the Sun on 28 January 2003. Its water outgassing rate probably
exceeded 10$^{30}$~\mols at that time (Povich et al. 2003). 
Solar elongation constraints prevented Odin observations before 2 March 2003. 
Odin monitored the rapid decrease of its activity while receding 
from the Sun. IRAM~30-m observations were done in parallel on the 12th 
of March 2003 (Biver et al. 2003). The spectra are shown in 
Fig.~\ref{02x5h2o} and show the rapid decrease of the line width and 
consequently coma expansion velocity. Production rates are
given in Table~\ref{tabqp} and suggest a rapid fall-off of the outgassing
rate with increasing heliocentric distance ($r_h$) in $r_h^{-3.6}$.

\begin{center} [Figure 3] \end{center}

\subsection{29P/Schwassmann-Wachmann 1}
	A deep integration on the H$_2$O line was done on this unusual 
distant comet in June 2003. Parallel observations of CO (26--29 June 2003) 
were obtained at IRAM-30m (Gunnarsson et al. 2004).
The comet had an outburst a few days before, around 14 June 2003 
at $m_1\approx12$ but was back to $m_1\approx13.5$ at the time of Odin 
observations. A marginal detection of H$_2$O was possibly obtained during 
the first series of orbits, but not confirmed later. Could this have been 
due to some residual outgassing of grains released by the earlier
outburst and that were dissipating? The inferred outgassing is below or 
comparable to that of CO ($Q_{CO}=4\times10^{28}$~\mols) -- assuming the 
same outgassing pattern from the nucleus and the same gas temperature (10~K).

\subsection{2P/Encke}
	Odin observed comet 2P/Encke at its favourable passage in 
November 2003, when it came within 0.19 AU from the Earth. 2P/Encke is the 
comet known to have
the shortest orbital period (3.3 years) and was then at its 59th observed 
return. The outgassing rate was one of the smallest measured in a comet
with Odin ($5-8\times10^{27}$~\mols), which yielded limited
signal-to-noise ratio in the maps. Fig.~\ref{h2oa22pk49p} shows the spectrum 
obtained on 23 November 2003 from the average of mapping points within a 
$2\times2$' box centred on the Odin beam. Ground-based observations were 
conducted in parallel to Odin observations (at CSO on the 16th and 
at IRAM~30-m on the 23rd, Biver et al. 2004).

\subsection{C/2001~Q4 (NEAT)}
	This dynamically new Oort cloud comet came to perihelion on 
16 May 2004 at 0.96 AU. It was the brightest comet of 2004 and reached a 
visual magnitude of $m_1=3.3$ in early May 2004. It has been extensively 
studied from the ground, especially starting in May 2004 since earlier 
it was a southern object. Odin monitored its outgassing rate in advance 
of these observing campaigns (Lecacheux et al. 2004).
This comet was bright enough to motivate searches of water isotopologues 
and ammonia around perigee (0.34 AU on 7 May) and to map the water emission. 
The H$_2^{18}$O line was detected with a signal-to-noise ratio of 13 and 
is shown in Fig.~\ref{q4h218o} together with the H$_2^{16}$O line
which was simultaneously observed with the AOS. The detection of NH$_3$ 
obtained at the same time with the third receiver is shown in 
Fig.~\ref{q4nh3}. Fig.~\ref{mapq4} shows
a plot of the intensity distribution of the water line on 16 May.

\begin{center} [Figure 4] \end{center}

\subsection{C/2002~T7 (LINEAR)}
	Comet C/2002~T7 (LINEAR) was the second bright comet of 2004, 
and also the target of an extensive ground-based observing campaign 
(with CSO, IRAM and many other facilities, Crovisier et al. 2005, 
Hatchell et al. 2005, DiSanti et al. 2004). 
A first distant perigee (1.56 AU) took place in the autumn of 2003 
when the comet was first detected in the radio (Crovisier et al. 2005). 
The comet was easily detected with Odin at the end of 
January 2004 (Table~\ref{tabqp}), still over 1.5 AU from the Sun. 
C/2002~T7 passed perihelion on 30 April 2004 at 0.62 AU from the Sun.
Its activity dropped relatively rapidly after it reached perigee at 0.27 AU on 
19 May 2004 but it was still bright enough to schedule Odin observations 
at the end of May, with searches of the two water isotopologues and ammonia: 
Fig.~\ref{t7h218o} shows the averages of
the H$_2^{18}$O and H$_2^{16}$O lines observed between 24 and 27 May 2005
and Fig.~\ref{t7nh3} shows the ammonia line detected in parallel with a 
signal-to-noise ratio of 5.

\begin{center} [Figure 5] \end{center}

\subsection{C/2003~K4 (LINEAR)}
	This comet was more active than anticipated when detected with the 
Nan\c{c}ay radio telescope in June 2004 with $Q_{\rm OH} > 10^{29}$~\mols. 
Odin observations were scheduled post-perihelion and took the relay of the 
Nan\c{c}ay monitoring of the water outgassing. Due to a signal stronger 
than anticipated, the Odin monitoring was extended to February 2005 and 
provided the most distant confirmed detection of the H$_2$O 557~GHz line 
in a comet, at 2.2 AU from the Sun, as shown
in Fig.~\ref{h2oa22pk49p}. The line recorded at $r_h=2.2$ AU is narrower 
($FWHM=0.95\pm0.08$ km~s$^{-1}$)
than the others (e.g. 2P/Encke and 9P/Tempel~1 which had a lower production 
rate though but were closer to the Sun), 
showing the decrease of expansion velocity with heliocentric distance.
Fig.~\ref{qp03k4} shows that we may have put into evidence the
turn-off of the water outgassing of the comet beyond 2 AU, where a more
rapid fall-off of the outgassing is clearly visible.

\subsection{C/2004~Q2 (Machholz)}
This long-period comet passed perihelion (1.21 AU) and perigee (0.35 AU) 
in January 2005. This was the fourth naked eye comet seen in less than a 
year and again a very promising observing target for ground-based radio 
investigations and for Odin. IRAM~30-m observations were conducted 
 on 14--18 January (Crovisier et al. 2005), at very close 
times to Odin ones (18--22 January), 
and provide a precise estimate of the gas temperature
($65\pm3$ K) from the observation of over 20 methanol lines.
H$_2^{16}$O and H$_2^{18}$O were clearly detected with Odin 
(Fig.~\ref{q2h218o}).
Ammonia was searched for in parallel with the AOS only, as the wide band is
necessary to monitor the slow drift in frequency of the ``572B1'' receiver, via
periodic observation of a telluric ozone line.
Following an unexpected shut-down probably due to a solar event, 
the AOS had to be switched off during a little more than half of the planned
NH$_3$ observations. 

\begin{center} [Figure 6] \end{center}

\subsection{9P/Tempel 1}
	Observing this comet in support to the Deep Impact mission 
(Meech et al. 2005) was a major objective of Odin in 2005. 
This  comet returned to perihelion on 5 July 2005 at 1.50 AU, the day 
after it was hit by the Deep Impact impactor (A'Hearn et al. 2005).
It is a 5.5-year orbital period comet belonging to the Jupiter-family group
of comets like 19P/Borrelly. 
95 Odin orbits were dedicated to the monitoring of 9P/Tempel~1 water 
outgassing rate between 18 June and 8 August 2005 (Biver et al. 2005). 
The data and results will be presented in a dedicated paper 
(Biver et al. 2006b).

\section{Comet maps}

Nine point maps, with typically 1' spacing, have been acquired 
on most comets in order to determine the position of the true centre of 
brightness.
The offset with respect to the expected position was generally less than 20"
and due to the limited accuracy of the telescope pointing (Section 2). 
Wider maps (7$\times$7' or more) have been obtained 
on bright comets C/2001~A2, C/2000~WM$_1$, 153P Lecacheux et al. (2003) and 
C/2001~Q4 (Fig.~\ref{mapq4}). 

Mapping the H$_2$O emission in cometary comae provides useful 
constraints for an accurate determination of the water production rate.
Indeed, the changes in intensity and velocity 
shift of the line with offset constrain the water
excitation mechanism and line optical thickness. As a first step
to analyze extended maps with our model (Section 4), we made radial 
averages of the signal and found a relatively good match of 
the evolution of line intensity and Doppler shift with distance to 
the nucleus, as illustrated in Fig.~\ref{profiliz}.

In addition, such maps aimed at providing information on the anisotropy 
of the outgassing. Some asymmetry was observed in the H$_2$O maps 
obtained for comet 19P/Borrelly (Bockel\'ee-Morvan et al. 2004). 
Asymmetric maps were expected
for comets 2P and 9P based on previous perihelion observations, but 
the signal-to-noise ratio was not high enough to retrieve significant 
information. Asymmetry seems marginally present in the extended maps of
C/2000~WM$_1$ and C/2001~Q4, as seen in Fig.~\ref{mapq4}. 

\begin{center} [Figure 9] \end{center}

\section{Water production rates}
	Line intensities have been converted into production rates 
(Table~\ref{tabqp}). A Haser model with symmetric outgassing 
and constant radial expansion velocity is used to describe the density, 
as in our previous studies (e.g. Biver et al. 1999). 
%It predicts a decrease of the gas 
%density as $1/r^2$ where $r$ is the distance to the nucleus multiplied 
%by an exponential decay factor due to photodissociation of water. 
The water photodissociation lifetime has been evaluated from the 
daily solar activity (Crovisier 1989). The expansion velocity $v_{exp}$ 
was generally inferred from the H$_2$O line shape 
(computation of line profiles predicts a half width at half maximum 
intensity on the red-shifted side about 0.1 km~s$^{-1}$ larger than 
$v_{exp}$), or from other contemporaneous observations of radio lines 
from CSO or IRAM~30-m.

Excitation of the water molecule rotational levels takes into account
collisions with neutrals at a constant gas temperature based on ground-based
measurements from other radio lines (e.g. Biver et al. 2006a) obtained 
nearly at the same time. Collisions with electrons also play a major role 
and are modelled according to Biver (1997) and Biver et al. (1999) with an 
electron density factor 
$x_{ne}$ set to 0.2 for all data. The corresponding electron density was
found to provide the best match to the radial evolution of line intensities
observed in extended maps, as illustrated in Fig.~\ref{profiliz}.  
The optical thickness of the water rotational lines is taken into account into 
the excitation process using the Sobolev ``escape probability'' 
method (Bockel\'ee-Morvan 1987).

	Radiation transfer takes into account line optical thickness.
The code can simulate line profiles that are generally
in good agreement with observed lines (e.g. Fig.~\ref{profiliz}), when the 
departure from isotropic outgassing is small. It is used to convert the line 
intensities into the water production rates, assuming isotropic outgassing.
An ortho-to-para ratio (OPR) of 3 has been assumed. 
Values of the OPR down to 2.4 (Crovisier et al. 1997) have been observed but
would only imply an underestimate of 6\% of the production rates given here.
In the case of isotropic outgassing, the modelling predicts a red-shift of the
line due to self-absorption in the foreground. For the strongest line observed
with peak intensities around 10~K, the Doppler shift predicted
($\sim+0.26$~\kms) is somewhat lower than observed (+0.28~\kms on average), 
suggesting that the opacity of the 557~GHz line, and so $Q_{\rm H_2O}$, 
may be slightly underestimated.
 
Evidence of anisotropy in the outgassing, often enhanced towards the Sun,
have been observed in several comets. Blue-shifted optically thin lines 
(H$_2^{18}$O, or HCN, CS, CH$_3$OH... from ground-based observations) 
are not uncommon. This is the case for comets C/2000~WM$_1$ 
(Biver et al. 2006a), 19P (Bockel\'ee-Morvan et al. 2004),
and C/2002~X5 around 12 March 2003, for which the optically thick
water line at 557~GHz appears less red-shifted than expected. For
C/2001~Q4 around 26 April to 2 May 2004 and C/2002~T7 between 24 and 
28 May 2004, we do not have simultaneous IRAM or CSO observations, 
but the H$_2^{18}$O line is blue-shifted and the red-shift of the 
H$_2^{16}$O is small, so that anisotropic outgassing is likely.

If the outgassing
rate is higher in the hemisphere facing the observer, then the total
water production rate is likely underestimated: self-absorption responsible
for the normal red-shift of the line is strongly auto absorbing the emission
from the hemisphere facing the observer so that increase of outgassing 
on this side will not increase much the emission. 
For example, in the case of C/2002~T7 on 26 May 2004, assuming that the 
outgassing takes place only on the observer facing hemisphere leads 
to $Q_{\rm H_2O}=32\times10^{28}$~\mols and a Doppler shift
$\delta v=-0.07$ km~s$^{-1}$. The isotropic assumption yields
$Q_{\rm H_2O}=22\times10^{28}$~\mols (for the same line intensity)
and $\delta v=+0.25$ km~s$^{-1}$: the observed Doppler shift 
($\delta v=0.0$ km~s$^{-1}$) is closer to the value expected for 
the asymmetric hypothesis.
Hence, production rates from Table~\ref{tabqp} must be used with 
caution, especially when anisotropic outgassing is suspected. 

\begin{center} [Figure 10] \end{center}

\section{Observations of the H$_2^{18}$O isotopologue}
	H$_2^{18}$O was first observed in comet 1P/Halley via mass spectroscopy
(Balsiger et al. 1995, Eberhardt et al. 1995).
Its first remote spectroscopic detection was obtained with Odin on comet 
153P/Ikeya-Zhang in 2003 (Lecacheux et al. 2003).
The $J_{Ka,Kc} = 1_{10}-1_{01}$ transition at 547.676~GHz has 
been since then securely detected in C/2001~Q4 (NEAT) 
(Fig.~\ref{q4h218o}), C/2002~T7 (LINEAR) (Fig.~\ref{t7h218o}) 
and C/2004~Q2 (Machholz) (Fig.~\ref{q2h218o}). From these spectra, 
we can readily see the difference between the optically thin line 
of H$_2^{18}$O and the optically thick H$_2^{16}$O line. Actually, the 
difference in velocity shifts of the lines is about 0.3~\kms~ in all cases. 

The $^{16}$O/$^{18}$O isotopic ratios in cometary water 
can be estimated from the $Q_{\rm H_2^{16}O}/Q_{\rm H_2^{18}O}$ ratio.
The following values are measured:
\begin{itemize}
\item Comet 153P: $^{16}$O/$^{18}$O = $530\pm60$ (Revised from Lecacheux et al. 2003);
\item Comet C/2001~Q4: $^{16}$O/$^{18}$O $\approx530\pm60$ %$479\pm48$
\item Comet C/2002~T7: $^{16}$O/$^{18}$O $\approx550\pm75$ %$464\pm63$
\item Comet C/2004~Q2: $^{16}$O/$^{18}$O = $508\pm33$
\end{itemize}
In the case of C/2001~Q4 and C/2002~T7 the H$_2^{16}$O production rates 
from Table~\ref{tabqp} have been multiplied by 1.1 and 1.2 to compensate 
the underestimation of $Q_{\rm H_2O}$ due to the non isotropic
outgassing (cf. Section 4). 

The case for the oxygen isotopic ratios in the Solar System is a
debated problem.  The terrestrial value (SMOW) for $^{16}$O/$^{18}$O
is 499.  Indeed, in primitive bodies such as carbonaceous chondrites,
most refractory inclusions show moderate anomalies in the 
$^{17}$O/$^{16}$O and $^{18}$O/$^{16}$O ratios 
($\delta ^{18}$O = (($^{18}$O/$^{16}$O)/($^{18}$O/$^{16}$O)$_{\rm SMOW}$-1)
= $-50$ to 0~\permil~, Krot et al. 2005).
Models of the pre-solar nebulae predict enrichments of $^{18}$O in 
cometary water up to $\delta ^{18}$O = 200~\permil~ (e.g., Yin 2004)

Our measurements in four comets ($\approx520$) are consistent 
with the terrestrial $^{16}$O/$^{18}$O ratio.
But imperfection in the modelling of the H$_2^{16}$O line 
makes the determination of the $^{16}$O/$^{18}$O ratio to be 
possibly slightly underestimated (Section 4).

\section{Observations of ammonia}
Due to the short NH$_3$ photodissociative lifetime (6700~s at 1 AU from 
the Sun), ammonia emission comes from the inner part of the atmosphere 
and requires a comet 
close to the Earth to be more easily detected. Several opportunities happened 
during Odin life and the $J_K = 1_0-0_0$ ammonia line at 572.498~GHz has been 
searched for in comets C/2001~Q4, C/2002~T7 and C/2004~Q2, which all came 
within 0.4 AU to the Earth.
Unfortunately, the phase lock of the ``572B1'' receiver did not 
work before 2003 and in 2005 when we attempted to search for NH$_3$ 
in comet C/2004~Q2 (Machholz).
It makes data reduction difficult as the frequency of the receiver 
slightly drifts with time and every spectrum needs to be carefully 
frequency calibrated
before addition. This is especially necessary for a narrow comet line.
Data obtained on comet C/2004~Q2 have not yet been reduced.
NH$_3$ was marginally detected in comets C/2001~Q4 and C/2002~T7 
(Figs.~\ref{q4nh3}, \ref{t7nh3}) with line areas of
$\int T_{mb}dv = 0.14\pm0.02$ K~km~s$^{-1}$ and $0.12\pm0.02$ K~km~s$^{-1}$, 
respectively.
Other ammonia lines are observable in the radio at 24~GHz, from the ground 
(e.g., Hatchell et al. 2005), but they are weaker and also suffer from beam dilution.

Production rates, assuming thermal equilibrium, are given in 
Table~\ref{tabqp} and abundances relative to water are:
\begin{itemize}
\item C/2001~Q4: NH$_3$/H$_2$O $= 0.50\pm0.09$\% ($Q_{\rm H_2O}=28\pm3\times10^{28}$)
\item C/2002~T7: NH$_3$/H$_2$O $= 0.33\pm0.08$\% ($Q_{\rm H_2O}=26\pm3\times10^{28}$)
\end{itemize}
Derived abundances are similar to values measured in other comets from 
direct observations of ammonia (Bird et al. 1999) or via observation 
of the NH$_2$ radical in the visible ($\sim0.5$\%, Kawakita and Watanabe 2002).
They are also consistent with the upper limits obtained in the same comets,
 from observations of the 24~GHz lines (Hatchell et al. 2005). 

\begin{center} [Figure 7, 8] \end{center}

\section{Conclusion}

In this paper, we presented preliminary results obtained on water and 
ammonia in comets using Odin:

\begin{itemize}
	\item Velocity resolved profiles of the optically thick 
557~GHz line were obtained: H$_2$O self-absorption is
observed and manifests itself as a red-shift of the line which varies 
throughout the coma.
	\item Variations of the gas expansion velocity with heliocentric 
distance and outgassing rate have been measured from the line widths.
	\item H$_2$O production rates were measured.
Though state-of-the-art excitation and radiative transfer models are used, some
uncertainties in production rate determinations (up to 50\%) remain due to the 
assumption of isotropic outgassing.
	\item The evolution of the water production rate with time and 
heliocentric distance has been monitored in several comets.
	\item The H$_2^{18}$O 547~GHz line was detected in four comets.
Comparing H$_2^{18}$O optically thin and H$_2^{16}$O optically thick 
lines provides constraints on the radiation transfer in the water dominated coma.
	\item With our current modelling of the H$_2^{16}$O line, we deduce a
$^{16}$O/$^{18}$O ratio around 520, compatible with, although marginally
higher than, the terrestrial value.
	\item Ammonia was detected at 572~GHz in two comets with abundance 
ratios ($\sim$0.4\%) similar to those measured in other comets.
\end{itemize} 
The current data set contains a wealth of information that will be used in 
the near future to improve the modelling of the
excitation of water in cometary comae. Detailed modelling, including 
anisotropic outgassing, is one of the next necessary steps to
retrieve accurate water production rates and explain the observed velocity 
shift of the water lines.

	In the near future the Herschel Space Observatory with its HIFI 
heterodyne instrument will allow to observe the full submillimetre spectrum 
of cometary water with high sensitivity and spatial resolution 
(Crovisier 2005). The MIRO experiment on board the Rosetta 
spacecraft (Gulkis et al. 2006) will observe in situ the same lines of H$_2$O 
isotopes and NH$_3$ in the coma of comet 67P/Churyumov-Gerasimenko and 
provide detailed spatial information. \\

{\it Acknowledgments.} Generous financial support from the research councils 
and space agencies in Canada, Finland France and Sweden is gratefully 
acknowledged. The authors thanks the whole Odin team, 
including the engineers who have been very supportive to
the difficult comet observations. Their help was essential in solving 
in near real time problems for such time critical observations. 
The help of Philippe Baron to retrieve, handle and reduce Odin data 
in Meudon was also appreciated. We also thank
Th\'er\`ese Encrenaz for presenting these results at the AOGS meeting
and inviting us to publish them in this issue.

\renewcommand{\baselinestretch}{1.0}

{\footnotesize
\begin{table}\vspace*{-1.5cm}
\begin{center}
\caption[]{Cometary molecular production rates}\label{tabqp}
\begin{tabular}{llccrll}
\hline
Comet & Mean UT date       & $<r_{h}>$ & $<\Delta>$ & \multicolumn{1}{c}{$Q_{\rm H_2O}$} & $Q_{\rm H_2^{18}O}$ &  $Q_{\rm NH_3}$  \\[0cm]
  & [yyyy/mm/dd.d] &  [AU] &  [AU] & [$10^{28}$~molec.~s$^{-1}$] & \multicolumn{2}{c}{[$10^{26}$~molec.~s$^{-1}$]} \\
\hline
C/2001~A2 (LINEAR) 
& 2001/04/27.3	& 0.938	& 0.826 & $10.5\pm1.3$ & & \\
& 2001/06/20.8	& 0.937 & 0.278 & $13.4\pm1.4$ & & \\
& 2001/06/21.7	& 0.947 & 0.273 & $11.2\pm1.2$ & & \\
& 2001/06/24.5	& 0.977 & 0.258 &  $8.9\pm0.7$ & & \\
& 2001/06/26.2	& 0.997 & 0.251 & $10.6\pm2.3$ & & \\
& 2001/06/28.4	& 1.023 & 0.245 &  $9.7\pm1.3$ & & \\
& 2001/06/30.4	& 1.047 & 0.244 &  $6.9\pm1.0$ & & \\
& 2001/07/01.8	& 1.062 & 0.244 &  $5.7\pm0.8$ & & \\
& 2001/07/02.5	& 1.072 & 0.245 &  $5.4\pm0.7$ & & \\
& 2001/07/03.0	& 1.078 & 0.246 &  $6.0\pm0.9$ & & \\
& 2001/07/06.1	& 1.117 & 0.256 &  $5.3\pm0.8$ & & \\
& 2001/07/07.6	& 1.135 & 0.264 &  $4.7\pm0.4$ & & \\
& 2001/07/09.3	& 1.157 & 0.273 &  $4.2\pm0.7$ & & \\
\hline
19P/Borrelly
& 2001/09/23.4	& 1.362	& 1.472 &  $3.8\pm0.5$ & & \\
& 2001/11/05.5	& 1.483	& 1.340 &  $2.6\pm0.4$ & & \\
\hline
C/2000~WM$_1$ (LINEAR)
& 2001/12/07.9  & 1.117 & 0.339 &  $5.0\pm0.8$ & & \\
& 2002/03/12.6  & 1.170 & 1.238 &  $6.7\pm0.5$ & & \\
\hline
153P/2002~C1 
& 2002/04/22.2  & 0.919 & 0.419 &  $33.4\pm1.4$ & & \\
\multicolumn{1}{r}{(Ikeya-Zhang)}
& 2002/04/24.6	& 0.961	& 0.415 &  $23.6\pm1.5$ & \vline &  \\
& 2002/04/27.1	& 0.997	& 0.407 &  $21.5\pm0.8$ & \vline ~$4.2\pm0.3$ &  \\
& 2002/04/28.2	& 1.022	& 0.405 &  $19.1\pm1.2$ & \vline & \\
\hline
C/2002 X5 
& 2003/03/03.4	& 0.987 & 0.934 &  $3.2\pm0.2$ & & \\
\multicolumn{1}{r}{(Kudo-Fujikawa)}
& 2003/03/12.5	& 1.178 & 1.092 &  $2.1\pm0.2$ & & \\
& 2003/03/21.7	& 1.361 & 1.304 &  $1.1\pm0.2$ & & \\
& 2003/03/30.6	& 1.529 & 1.540 &  $0.8\pm0.3$ & & \\
\hline
29P/Schwassmann-W. 1
& 2003/06/26.5	& 5.752 & 5.305	&   $<2.5\;\;\;$ & & \\
\hline
2P/Encke
& 2003/11/16.7	& 1.014 & 0.261 &  $0.49\pm0.07$ & & \\
& 2003/11/23.7	& 0.896 & 0.275 &  $0.78\pm0.16$ & & \\
\hline
C/2001~Q4 (NEAT)
& 2004/03/06.6  & 1.518 & 1.734 & $17.1\pm0.4$ & & \\
& 2004/03/15.6  & 1.413 & 1.534 & $16.3\pm0.7$ & & \\
& 2004/03/24.6  & 1.311 & 1.319 & $17.4\pm1.0$ & & \\
& 2004/04/02.6  & 1.216 & 1.087 & $20.0\pm1.4$ & & \\
& 2004/04/13.6  & 1.113 & 0.790 & $26.0\pm4.0$ & & \\
& 2004/04/29.8  & 1.000 & 0.383 & $^a25.4\pm1.7$ & $5.3\pm0.4$ & $14.1\pm2.1$ \\
& 2004/05/16.0	& 0.962 & 0.436 & $22.2\pm2.9$ & & \\
\hline
C/2002~T7 (LINEAR)
& 2004/01/26.6    & 1.757 & 1.860 & $32.4\pm1.1$ & & \\
& 2004/02/01.7    & 1.666 & 1.910 & $28.0\pm0.9$ & & \\
& 2004/05/25.9    & 0.931 & 0.407 & $^a21.8\pm2.3$ & $4.7\pm0.4$ & $8.7\pm1.7$ \\
& 2004/05/29.2    & 0.973 & 0.488 & $19.2\pm2.9$ &$2.5\pm0.7$ & \\
\hline
C/2003~K4 (LINEAR)
& 2004/11/27.7    & 1.269 & 1.395 & $21.8\pm1.5$ & & \\
& 2004/12/15.6    & 1.453 & 1.181 & $17.0\pm0.3$ & &  \\
& 2004/12/28.7    & 1.601 & 1.164 & $14.5\pm0.6$ & &  \\
& 2005/01/05.7    & 1.694 & 1.229 & $12.8\pm0.9$ & &  \\
& 2005/01/17.7    & 1.836 & 1.419 & $11.7\pm0.4$ & &  \\
& 2005/02/04.7    & 2.053 & 1.842 &  $6.6\pm0.5$ & &  \\
& 2005/02/19.7    & 2.233 & 2.247 &  $3.8\pm0.4$ & &  \\
\hline
C/2004~Q2 (Machholz)
& 2005/01/20.0    & 1.208 & 0.396 & $26.4\pm0.8$ & $5.2\pm0.2$ & \\
\hline
9P/Tempel 1
& 2005/06/18.2	& 1.516 & 0.818 &  $1.15\pm0.10$ & &  \\
& 2005/06/23.7	& 1.511 & 0.842 &  $1.24\pm0.13$ & &  \\
\hline
\end{tabular}
\end{center}
$^a$: Value probably underestimated by about 10--20\% due to asymmetric outgassing (see text).
The uncertainty on the production rate takes into account dispersion from 
values determined from different offset points when available.
\end{table}
}

\newpage
\begin{center}

\begin{figure*}\vspace{0cm}\hspace{1cm}
   \psfig{width=18cm,angle=270,figure=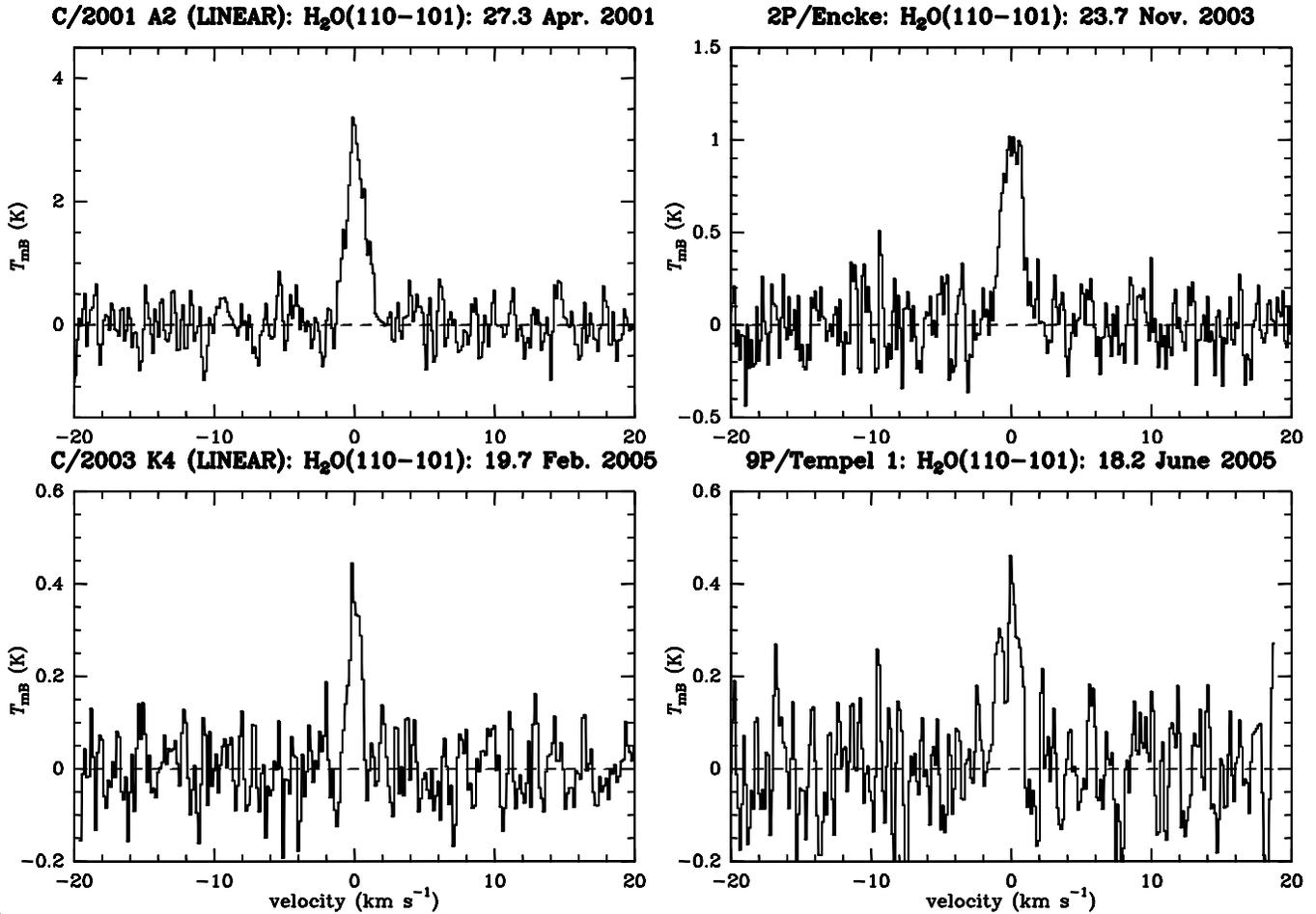}\vspace{-0.5cm}
   \caption{Sample of observations of the 556.9~GHz H$_2$O line with Odin: 
    C/2001 A2 (LINEAR) on 27 April 2001
    (first observation, mean pointing offset is 39''), 2P/Encke on 23 November 2003
    (average of centred and offset positions up to 1.5': mean offset is 51''), 
    C/2003 K4 (LINEAR) on 19 February 2005 (most distant detection,
    $r_h$=2.23 AU, centred position) and 9P/Tempel 1 on 18 June 2005 
    (first observation of the Deep Impact campaign, % with only one receiver and 
    average of centre and 1' offset positions: mean offset is 46''). 
    The vertical scale is the intensity
    in main beam brightness temperature and the horizontal scale is the Doppler
    velocity with respect to the comet nucleus.}
   \label{h2oa22pk49p}
\end{figure*}\vspace{0cm}

\begin{figure*}\vspace{-2cm}\hspace{1cm}
   \psfig{width=16cm,angle=270,figure= 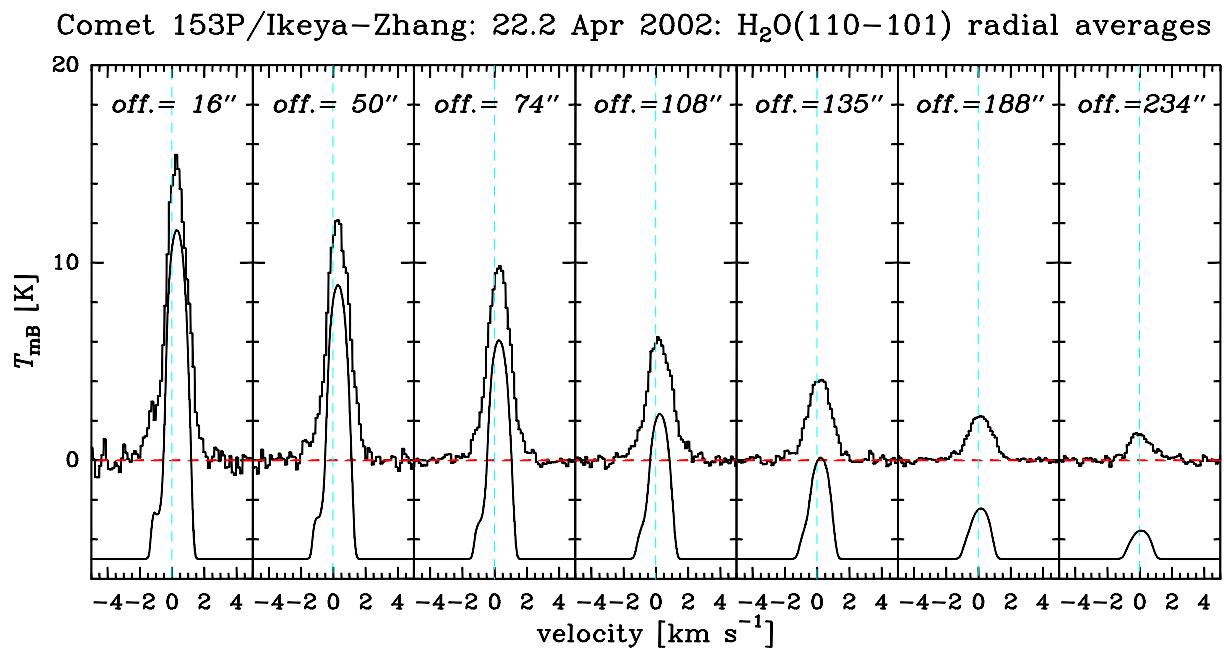}\vspace{-0.5cm}
   \caption{Observed (top) and simulated (shifted down by 5 K) H$_2$O line profiles 
   versus radial distance from the nucleus in the coma of comet 153P/Ikeya-Zhang. 
   Data are extracted from the radial
   averages of the map obtained on 22 April 2002.
   The differences of the line areas and Doppler shifts between predicted and
   observed values are below 3-$\sigma$.}
   \label{profiliz}
\end{figure*}\vspace{-6cm}

\begin{figure*}\vspace{-4cm}\hspace{1cm}
   \psfig{width=16cm,angle=270,figure=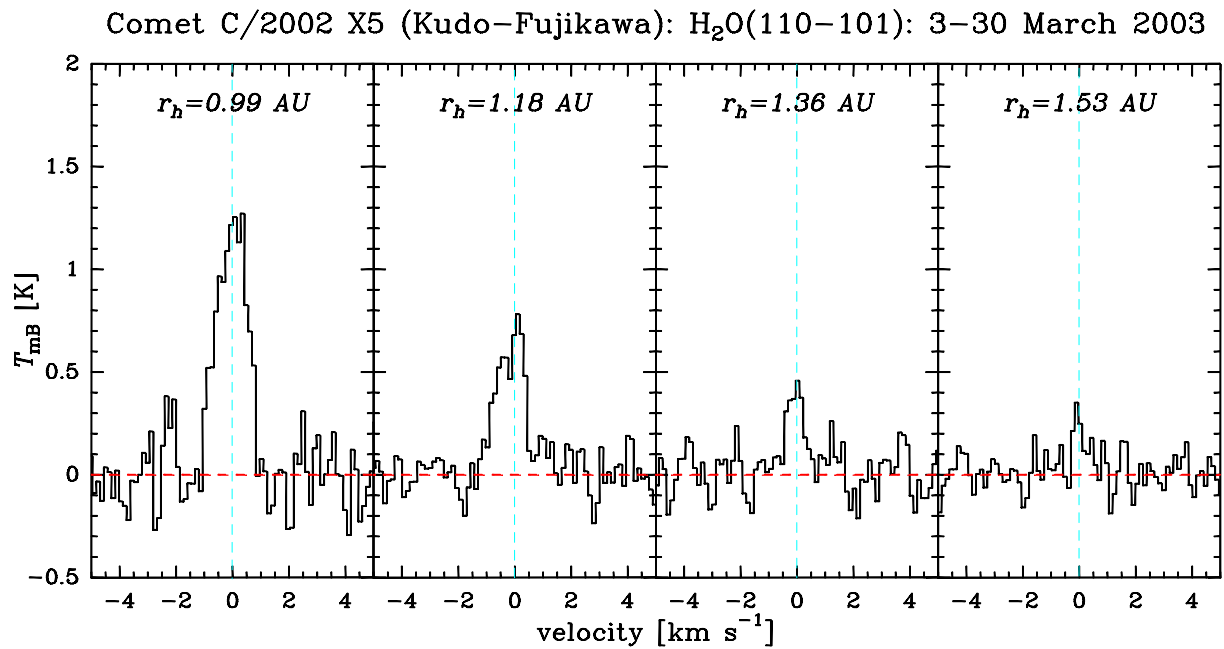}\vspace{-0.5cm}
   \caption{H$_2$O($1_{10}-1_{01}$) line at 556.9~GHz observed by Odin in
   comet C/2002~X5 (Kudo-Fujikawa) on 3, 12, 21 and 30 March 2003.
   Note the decrease of the line intensity and width as the
   comet is moving away from the Sun and decreasing in activity.}
   \label{02x5h2o}
\end{figure*}

\begin{figure*}\hspace{2cm}
   \vbox{\psfig{figure=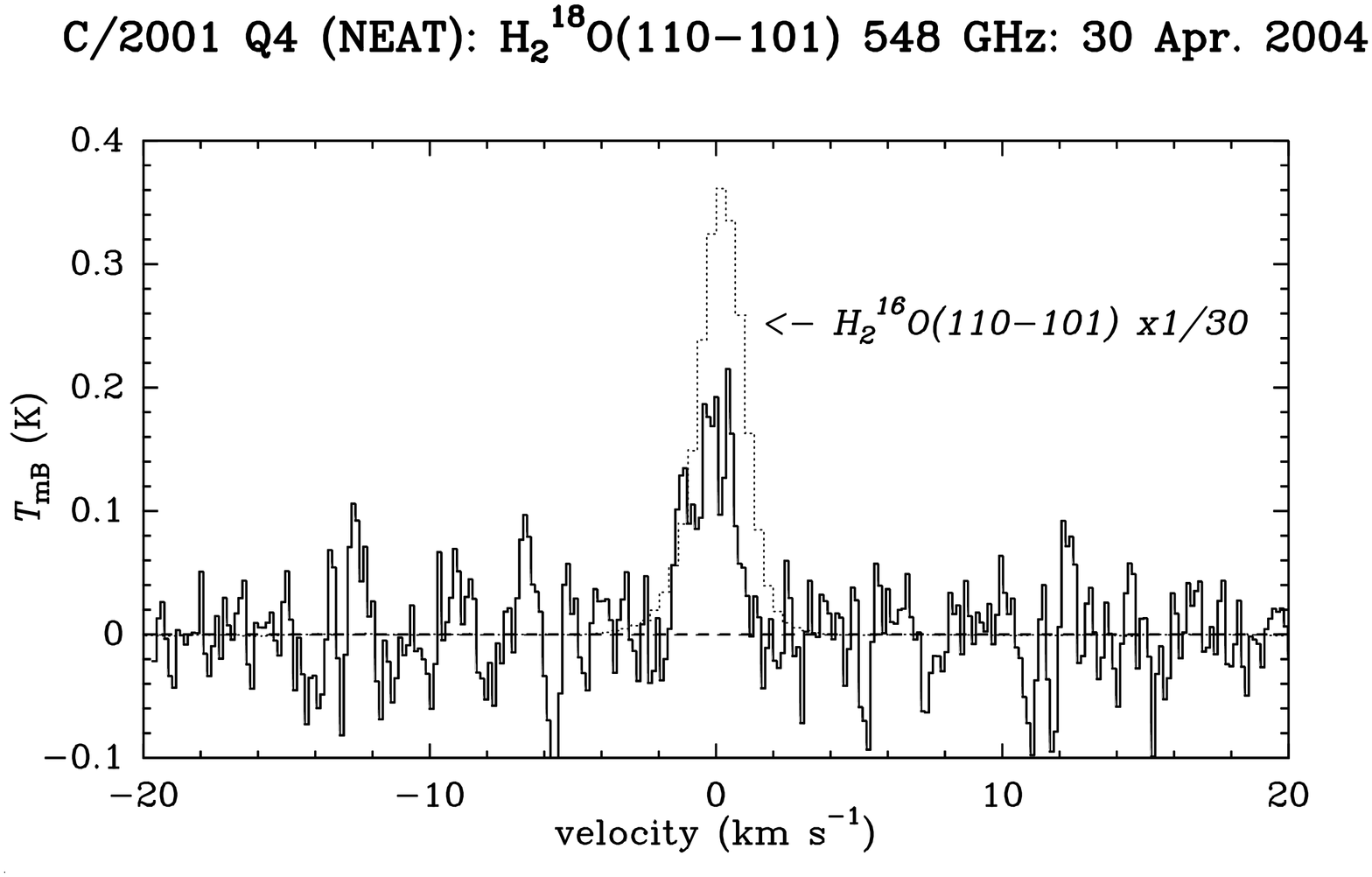,width=14cm,angle=270}}\vspace{-0.5cm}
   \caption{The 547.7~GHz H$_2^{18}$O line observed with Odin and the AC2 autocorrelator
    in comet C/2001~Q4 (NEAT): average of 26.7 April to 2.9 May data.
    Dotted line: the  556.9~GHz H$_2^{16}$O line observed simultaneously  
    with the acousto optical spectrometer, scaled down by a factor of 30.}
   \label{q4h218o}
   \end{figure*}

\begin{figure*}\hspace{2cm}
   \psfig{width=14cm,angle=270,figure=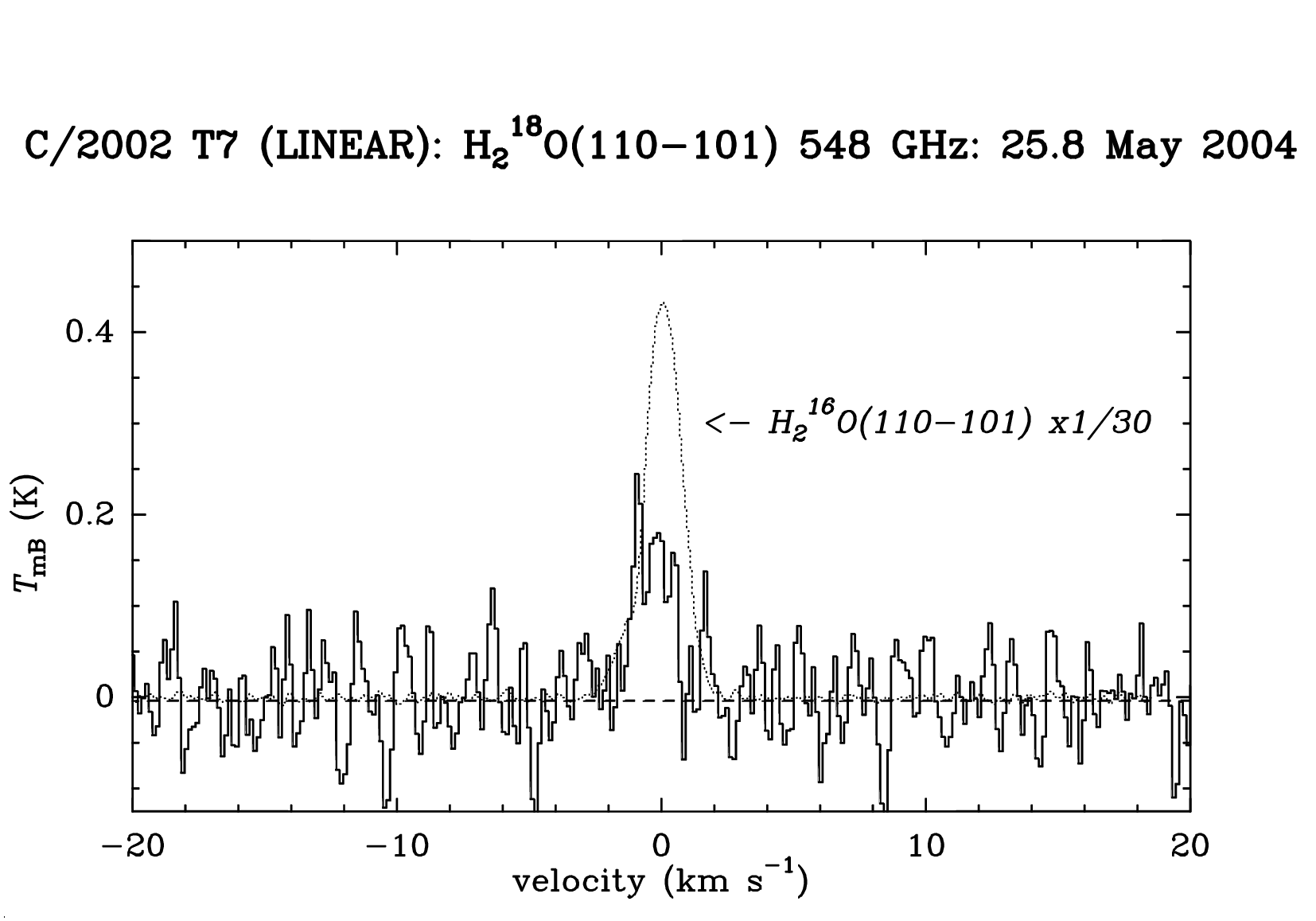}\vspace{-0.5cm}
   \caption{The 547.7~GHz H$_2^{18}$O line observed with Odin and the AC2 autocorrelator
    in comet C/2002 T7 (LINEAR): average of 24.1 to 27.5 May data.
    Dotted line: the  556.9~GHz H$_2^{16}$O line observed during the same 
    time interval, scaled down by a factor of 30.}
   \label{t7h218o}
\end{figure*}

\begin{figure*}\hspace{2cm}
   \psfig{width=14cm,angle=270,figure=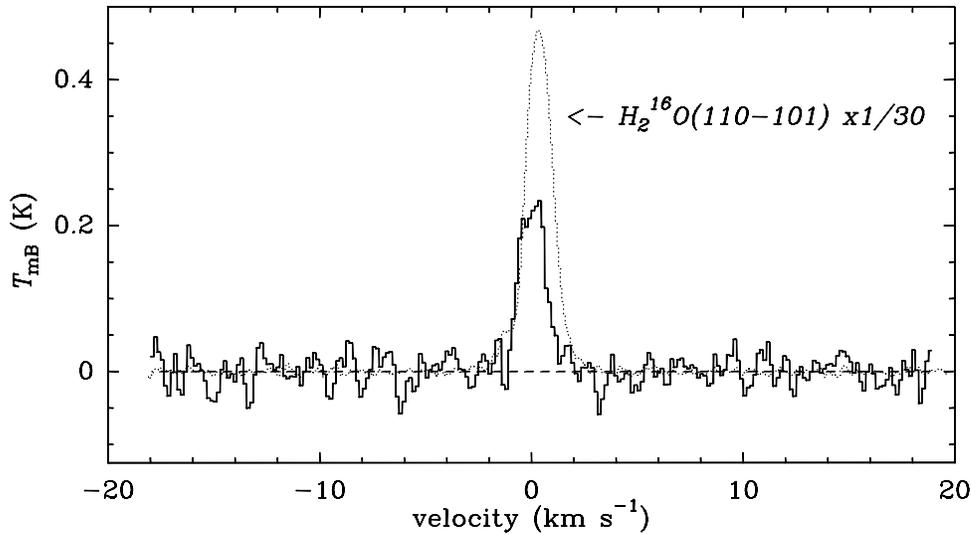}\vspace{-0.5cm}
   \caption{The 547.7~GHz H$_2^{18}$O line observed with Odin and both autocorrelators
    in comet C/2004~Q2 (Machholz): average of 17.8 to 23.8 January data.
    Dotted line: the  556.9~GHz H$_2^{16}$O line observed during similar
    time interval (17.8--21.8 Jan.), scaled down by a factor of 30.}
   \label{q2h218o}
   \end{figure*}

\begin{figure*}\hspace{2cm}
   \psfig{width=14cm,angle=270,figure=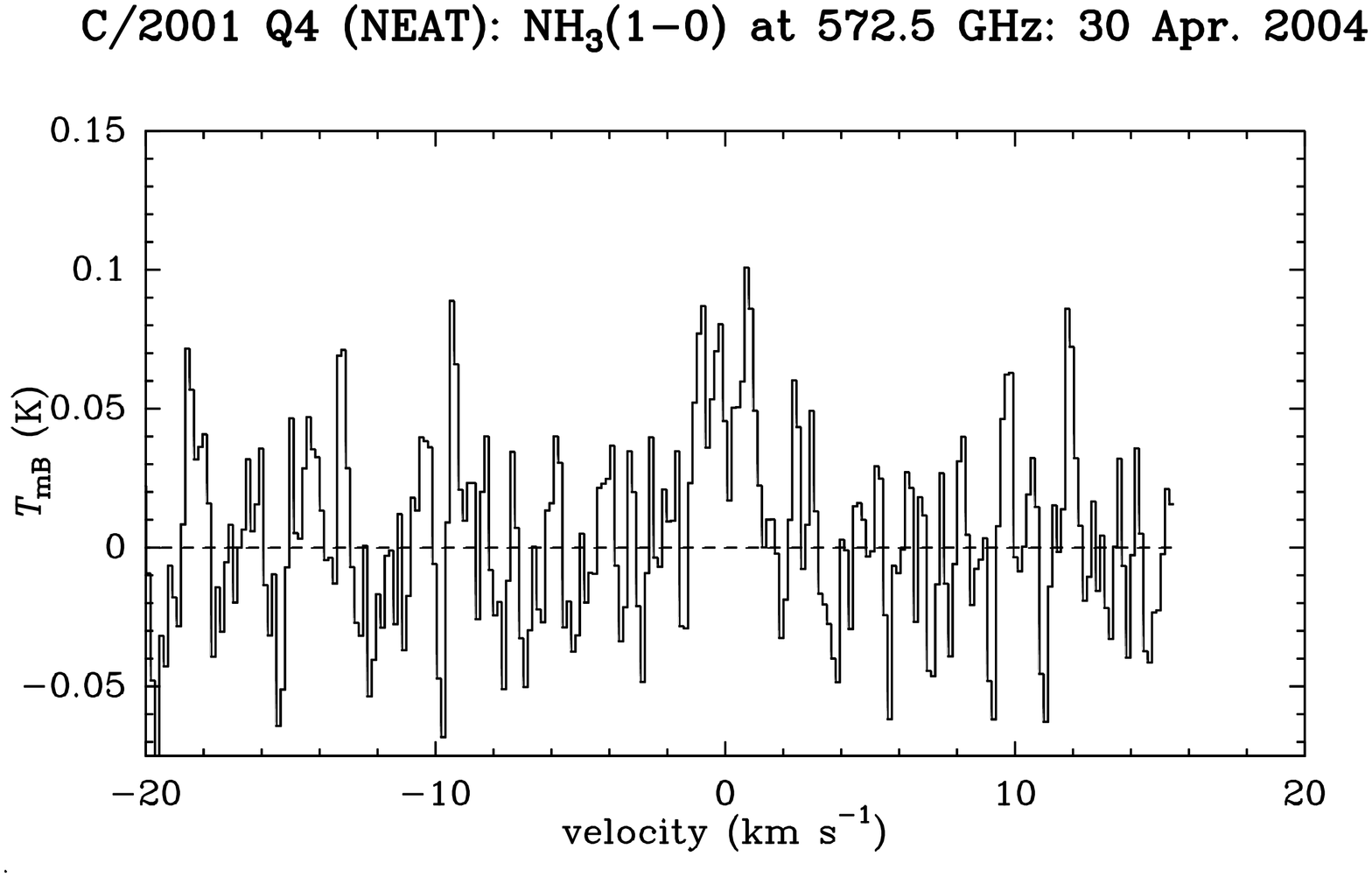}\vspace{-0.5cm}
   \caption{The 572.5~GHz NH$_3$ line observed with Odin and the AC1 autocorrelator
    in comet C/2001 Q4 (NEAT): average of 26.7 April to 2.9 May data.
    H$_2^{16}$O and H$_2^{18}$O observations were conducted in parallel.}
   \label{q4nh3}
   \end{figure*}

\begin{figure*}\vspace{4cm}\hspace{2cm}
   \psfig{width=14cm,angle=270,figure=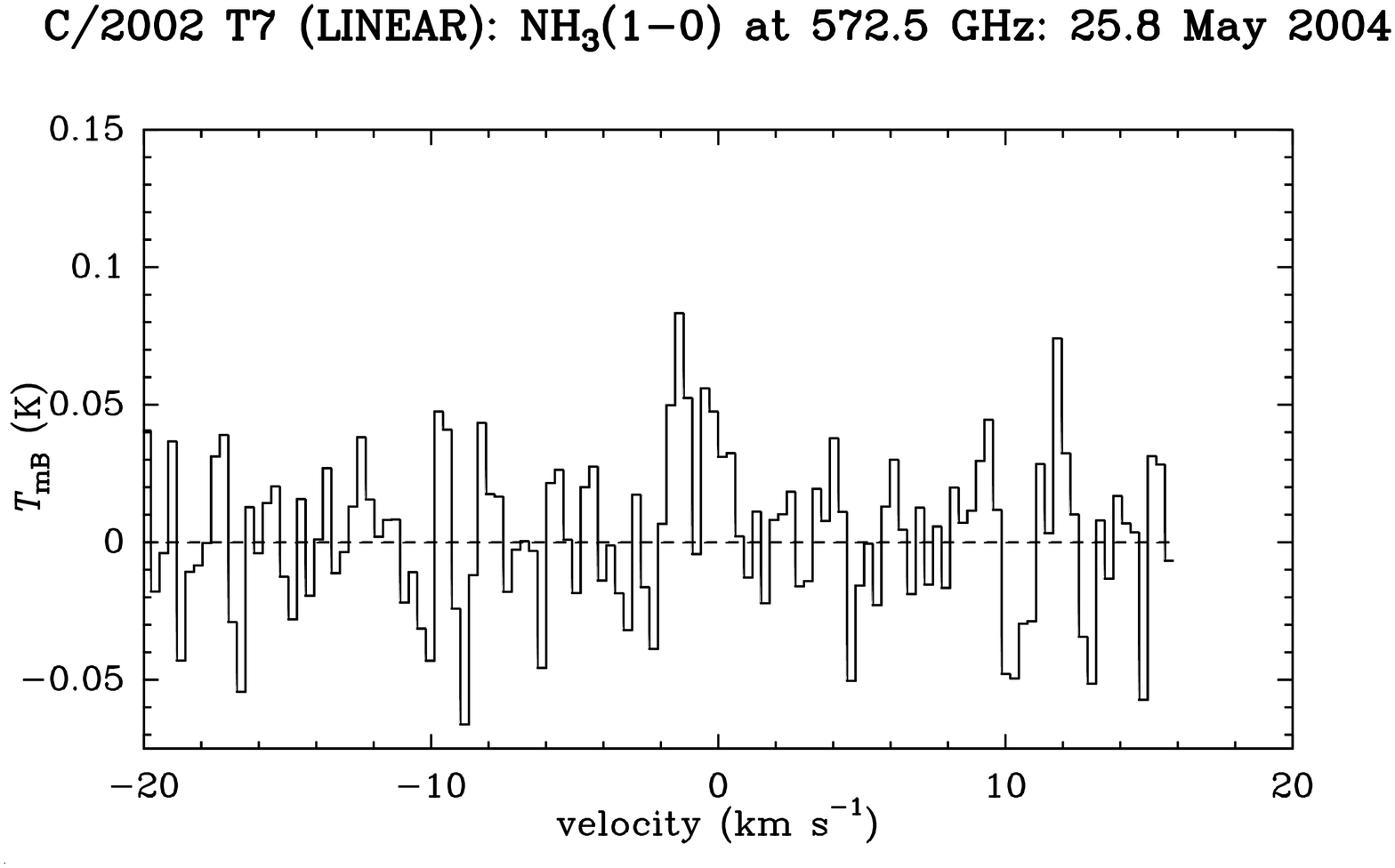}\vspace{-0.5cm}
   \caption{The 572.5~GHz NH$_3$ line observed with Odin and the AC1 autocorrelator
    in comet C/2002~T7 (LINEAR): average of 24.1 to 27.5 May data.
    Due to the lack of terrestrial line at this frequency, the velocity scale
    has yet to be calibrated and can be still off by up to 0.5~km/s. But 
    as for C/2001~Q4 observations (Fig.~\ref{q4h218o}, \ref{q4nh3}), comparison to 
    H$_2^{18}$O line observed simultaneously (Fig.~\ref{t7h218o}) suggests that the 
    blue-shift of the line is real.}
   \label{t7nh3}
   \end{figure*}

\begin{figure*}\vspace{-1.5cm}\hspace{2cm}
   \psfig{width=12cm,angle=0,figure=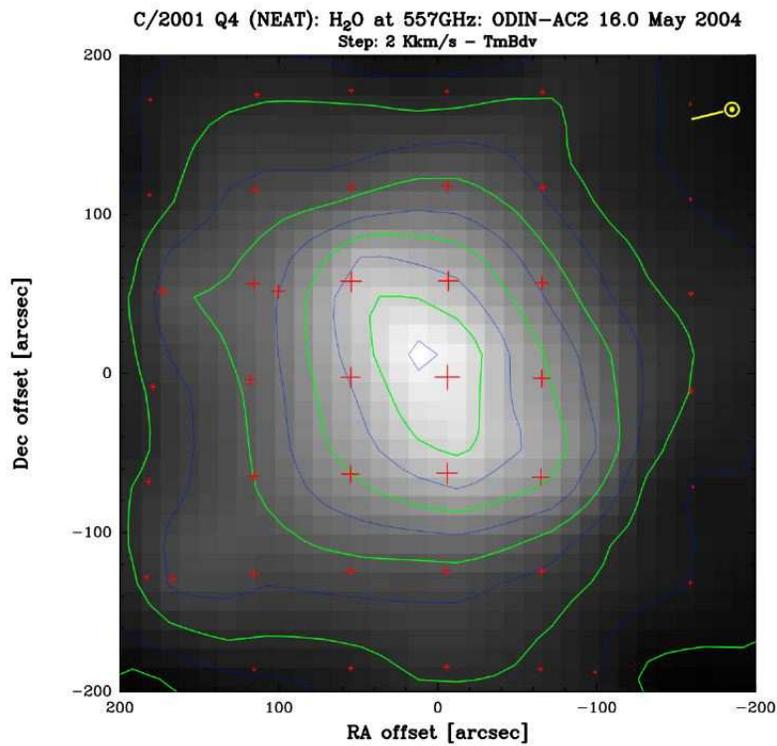}\vspace{-0.5cm}
   \caption{Gray scale and contours of the integrated intensity of the 
   556.9~GHz water line mapped, with a 7$\times$7 point grid at 1' spacing, 
   in comet C/2001 Q4 (NEAT) on 16 May 2004.
   Crosses correspond to actual measurements, with size proportional 
   to line integrated intensity. The peak intensity 
   is 19 K~km~s$^{-1}$ and contours are drawn by steps of 2 K~km~s$^{-1}$.
   Some extension perpendicular to the Solar direction indicated in the upper right
   seems to be present. The phase angle was 84$\deg$ during the observations.}
   \label{mapq4}
   \end{figure*}\vspace{-4cm}

\begin{figure*}\vspace{-1cm}\hspace{1cm}
   \psfig{width=14cm,angle=270,figure=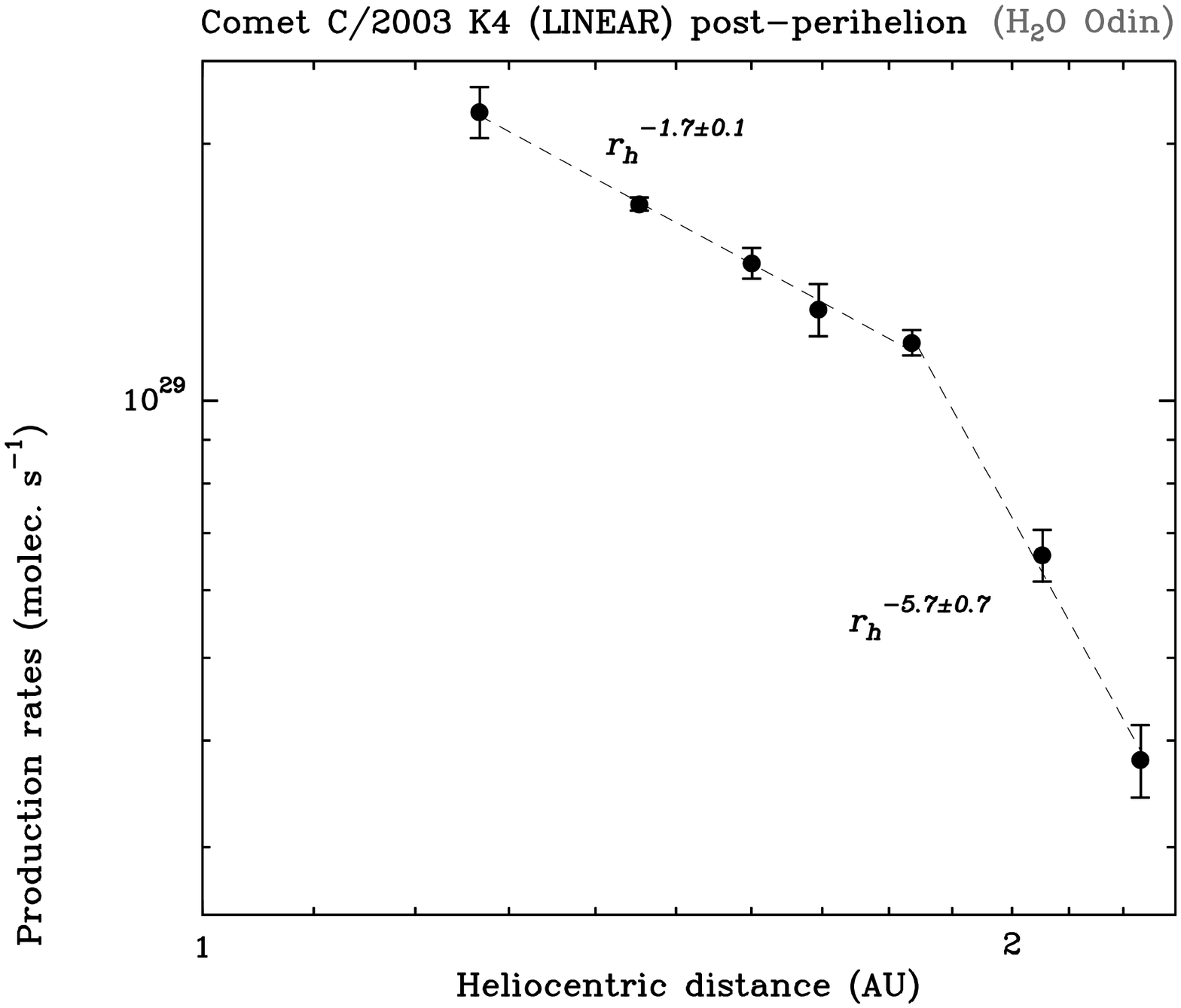}\vspace{-0.5cm}
   \caption{The evolution of the water outgassing rate of comet C/2003~K4 (LINEAR)
   with heliocentric distance when the comet receded from the Sun between Nov. 2004 
  and Feb. 2005. Note the change of slope around 1.9 AU.}
   \label{qp03k4}
   \end{figure*}

\end{center}


\begin{thebibliography}{}
\bibitem[A'Hearn et al. 2005]{Ahe05}
	A'Hearn, M. F., Belton, M. J. S., Delamere, W. A., 2005,
	Deep Impact: excavating comet Tempel 1,
	{\it Science}, 310, 258--264 
\bibitem[Balsiger et al. 1995]{Bal95}
	Balsiger, H., Altwegg, K. and Geiss, J., 1995,
	D/H and $^{18}$O/$^{16}$O ratio in the hydronium ion and in neutral 
	water from in situ ion measurements in comet Halley,
	{\it J. Geophys. Res.}, 100, 5827--5834
\bibitem[Bird et al. 1999]{Bir99}
	Bird, M. K., Janardhan, P., Wilson, T. L. et al., 1999,
	K-band radio observations of comet Hale-Bopp: detections of 
	ammonia and (possibly) water,
	{\it Earth, Moon, and Planets}, 78, 21--28 
\bibitem[Biver (1997)]{Biv97}
        Biver, N., 1997, 
	Mol\'ecules m\`eres com\'etaires: observations et mod\'elisations.
	{\it Ph.D. Thesis, Universit\'e Paris VII}
\bibitem[Biver et al. (1999)]{Biv99}
        Biver, N., Bockel\'ee-Morvan, D., Crovisier, J., et al., 1999, 
	Spectroscopic monitoring of comet C/1996 B2 (Hyakutake) with the 
	JCMT and IRAM radio telescopes, 
	{\it A.J.}, 118, 1850--1872
\bibitem[Biver et al. 2003]{Biv03}
        Biver, N., Bockel\'ee-Morvan, D., Crovisier, J., et al., 2003,
	Radio observations of comets C/2002 X5 (Kudo-Fujikawa) and 
	C/2002 V1 (NEAT) at very small heliocentric distances,
	{\it Bulletin of the AAS}, 35, 968
\bibitem[Biver et al. 2004]{Biv04}
	Biver, N., Bockelée-Morvan, D., Colom, P., et al., 2004,
	Recent chemical investigation of comets from radio spectroscopy,
	{\it 35th COSPAR Scientific Assembly, 18--25 July 2004}, p. 2772
\bibitem[Biver et al. 2005]{Biv05}
	Biver, N.; Bockelée-Morvan, D., Boissier, J., et al. 2005,
	Radio observations of comet 9P/Tempel 1 before and after Deep Impact,
	{\it Bulletin of the AAS}, 37, 710	
\bibitem[Biver et al. 2006a]{Biv06}
        Biver, N., Bockel\'ee-Morvan, D., Crovisier, J., et al., 2006a, 
	Radio wavelength molecular observations of comets C/1999~T1 (McNaught-Hartley), 
	C/2001~A2 (LINEAR), C/2000~WM$_1$ (LINEAR) and 153P/Ikeya-Zhang,
	{\it A\&A}, 449, 1255--1270
\bibitem[Biver et al. 2006b]{Biv06b}
        Biver, N., Bockel\'ee-Morvan, D., Crovisier, J., et al., 2006b,
	Radio observations of comet 9P/Tempel 1 before and after Deep Impact, 
	{\it Icarus}, {\it in press}
\bibitem[Bockel\'ee-Morvan 1987]{Boc87}
        Bockel\'ee-Morvan, D., 1987, 
	A model for the excitation of water in comets, 
	{\it A\&A}, 181, 169--181
\bibitem[Bockel\'ee-Morvan et al. 2004]{Boc04}
        Bockel\'ee-Morvan, D., Biver, N., Colom, P., et al., 2004, 
	The outgassing and composition of comet 19P/Borrelly 
	from radio observations, 
	{\it Icarus}, 167, 113--128
\bibitem[Colom et al. 2004]{Col04}
        Colom, P., Biver, N., Lecacheux, A., Crovisier, J. 
	and Bockel\'ee-Morvan, D., 2004,
	Water outgassing rates in comets with ODIN and 
	the Nan\c{c}ay radio telescope,
	In: Combes, F. et al. (Ed.), {\it SF2A Scientific Highlights 2004},
	EdP-Sciences Conference Series, 69--70
\bibitem[Crovisier 1989]{Cro89} 
        Crovisier, J., 1989, 
	The photodissociation of water in cometary atmospheres, 
	{\it A\&A}, 213, 459--464
\bibitem[Crovisier et al. 1997]{Cro97}
        Crovisier, J., Leech, K., Bockel\'ee-Morvan, D., et al., 1997,
	The spectrum of comet Hale-Bopp (C/1995 01) observed with the 
	Infrared Space Observatory at 2.9 AU from the Sun
	{\it Science}, 275, 1904--1907
\bibitem[Crovisier et al. 2005]{Cro05}
        Crovisier, J., Biver, N., Bockel\'ee-Morvan, D., et al., 2005,
	Chemical diversity of comets observed at radio wavelengths in 2003-2005,
	{\it Bulletin of the AAS}, 37, 646
\bibitem[Crovisier 2005]{Cro05b}
        Crovisier, J., 2005, Comets and asteroids with the Herschel Space 
	Observatory, In: Wilson, A. (Ed.), {\it Proceedings of the dusty and 
	molecular universe: a prelude to Herschel and ALMA}, 
	ESA SP-577, 145--150 
\bibitem[DiSanti et al. 2004]{Dis04}
	DiSanti, M. A., Reuter, D. C., Mumma, M. J., et al., 2004,
	Modeling formaldehyde $\nu_1$ and $\nu_5$ band emission in 
	comet C/2002 T7 (LINEAR), {\it Bulletin of the AAS}, 36, 23.08
\bibitem[Eberhardt et al. 1995]{Ebe95}
	Eberhardt, P., Reber, M., Krankowsky, D. and Hodges, R. R., 1995,
	The D/H and $^{18}$O/$^{16}$O ratios in water from comet P/Halley,
	{\it A\&A}, 302, 301--316
\bibitem[Frisk et al. 2003]{Fri03}  
        Frisk, U., Hagstr\"om, M., Ala-Laurinaho, J. et al., 2003, 
	The Odin satellite. I. Radiometer design and test,
	{\it A\&A}, 402, L27--L34
\bibitem[Gulkis et al. 2006]{Gul06}
        Gulkis, S., Allen, M., Backus, C., et al., 2006, 
	Remote sensing of a comet at millimeter and submillimeter 
	wavelengths from an orbiting spacecraft, 
	{\it Planet. Space Sci.}, {\it this issue}
\bibitem[Gunnarsson et al. 2004]{Gun04}
	Gunnarsson, M., Biver, N., Bockel\'ee-Morvan, D., Crovisier, J., 
	Rickman, H. and Festou, M., 2004,
	Mapping the CO coma of comet 29P/Schwassmann-Wachmann 1,
	{\it 35th COSPAR Scientific Assembly, 18--25 July 2004}, p. 2872
\bibitem[Hatchell et al. 2005]{Hat05} 
	Hatchell, J., Bird, M. K., van der Tak, F. F. S. and Sherwood, W. A., 2005, 
	Recent searches for the radio lines of NH$_3$ in comets,
	{\it A\&A}, 439, 777--784
\bibitem[Hjalmarson et al. 2003]{Hja03}
        Hjalmarson, \AA., Frisk, U., Olberg, M. et al., 2003, 
	Highlights from the first year of Odin observations, 
	{\it A\&A}, 402, L39--L46
\bibitem[Hjalmarson et al. (2005)]{Hja05}
        Hjalmarson, \AA., Bergman, P., Biver, N. et al., 2005, 
	Recent astronomy highlights from the Odin satellite,
	{\it Advances in Space Research}, 36, 1031--1047
\bibitem[Kawakita and Watanabe 2002]{Kaw02}
	Kawakita, H. and Watanabe, J., 2002, 
	Revised fluorescence efficiencies of cometary NH$_2$: 
	ammonia abundance in comets, 
	{\it ApJ}, 572, L177--L180
\bibitem[Krot et al. 2005]{Kro05}
	Krot, A.N., Hutcheon, I.D., Yurimoto, H. et al., 2005, 
	Evolution of oxygen isotopic composition in the inner Solar Nebula, 
	{\it ApJ}, 622, 1333--1342
\bibitem[Lecacheux et al. (2003)]{Lec03} 
        Lecacheux, A., Biver, N., Crovisier, J. et al., 2003, 
	Observations of water in comets with Odin, 
	{\it A\&A}, 402, L55--L58
\bibitem[Lecacheux et al. 2004]{Lec04}
	Lecacheux, A., Biver, N., Crovisier, J. and Bockel\'ee-Morvan, D., 2004,  
	Comet C/2001 Q4 (NEAT), 
	{\it IAU Circular } No 8304
\bibitem[Meech et al. 2005]{Mee05}
	Meech, K. J., A'Hearn, M. F., Fernández, Y. R., et al., 2005,
	The Deep Impact earth-based campaign,
	{\it Space Science Reviews}, 117, 297--334 
\bibitem[Neufeld et al. 2000]{Neu00} 
	Neufeld, D.A., Stauffer, J.R., Bergin, E.A. et al., 2000,
	Submillimeter Wave Astronomy Satellite observations of 
	water vapor toward comet C/1999 H1 (Lee), 
	{\it ApJ}, 539, L151--L154
\bibitem[Nordh et al. 2003]{Nor03} 
	Nordh, H.L., von Sch\'eele, F., Frisk, U. et al., 2003, 
	The Odin orbital observatory, 
	{\it A\&A}, 402, L21--L25
\bibitem[Povich et al. 2003]{Pov03} 
	Povich, M.S., Raymond, J.C., Jones, G.H. et al., 2003, 
	Doubly ionized carbon observed in the plasma tail of comet 
	Kudo-Fujikawa, 
	{\it Science}, 302, 1949--1952
\bibitem[Yin 2004]{Yin04}
	Yin, Q.-Z., 2004, Predicting the Sun's oxygen isotope composition,
	{\it Science}, 305, 1729--1730

\end{thebibliography}
\end{document}